\documentclass[aps,pre,twocolumn,float,showpacs,amsmath,amssymb,floatfix, superscriptaddress]{revtex4}
\usepackage{graphicx}
\usepackage{color}

\begin{document}              

\title{Random fluctuation leads to forbidden escape of particles \\ Phys. Rev. E 82 026211 (2010)}

\author{Christian
S. Rodrigues\footnote{Electronic Address: christian.rodrigues@mis.mpg.de}}
\affiliation{Max Planck Institute for Mathematics in the Sciences, Inselstr., 22, 04103 Leipzig, Germany}
\affiliation{Department of Physics and Institute for Complex Systems and Mathematical Biology,
King's College, University of Aberdeen - Aberdeen AB24 3UE, UK}

\author{Alessandro P. S. de Moura}
\affiliation{Department of Physics and Institute for Complex Systems and Mathematical Biology,
King's College, University of Aberdeen - Aberdeen AB24 3UE, UK}

\author{Celso Grebogi} 
\affiliation{Department of Physics and Institute for Complex Systems and Mathematical Biology,
  King's College, University of Aberdeen - Aberdeen AB24 3UE, UK}

\date{\today}

\begin{abstract}


A great number of physical processes are described within the context
of Hamiltonian scattering. Previous studies have rather been focused on
trajectories starting outside invariant structures, since the ones
starting inside are expected to stay trapped there forever. This is
true though only for the deterministic case. We show however that,
under finitely small random fluctuations of the field, trajectories
starting inside Arnold-Kolmogorov-Moser (KAM) islands escape within finite time. The
non-hyperbolic dynamics gains then hyperbolic characteristics due to
the effect of the random perturbed field. As a consequence,
trajectories which are started inside KAM curves escape with
hyperbolic-like time decay distribution, and the fractal dimension of a 
set of particles that remain in the scattering region approaches that for 
hyperbolic systems. We show a universal quadratic
power law relating the exponential decay to the amplitude of noise. We
present a random walk model to relate this distribution to the
amplitude of noise, and investigate this phenomena with a numerical
study applying random maps.

\end{abstract}


\maketitle



\indent


Scattering is the general term used to describe systems whose dynamics take place in an unbounded phase space, and such that the dynamics is trivial outside of a localised region called the \emph{interaction region}.  A classical example is a thrown particle which collides or interacts with a fixed target, then escapes from the neighbourhood of the obstacle. If the region where such particle interacts with the obstacle is much smaller than the whole space we locally assume that the particle comes from infinity and is scattered again towards infinity. Alternatively, we may want to consider the dynamics of initial conditions starting inside the scattering region, in other words, the dynamics of an open system.  We represent it by the dynamics, $x_{n+1} = f(x_{n})$, taking place in an unbounded phase space $M$ such that, the scattering region is $\textbf{W} \subset M$. Accordingly, the orbits of initial conditions in $\textbf{W}$, the region of interest, may eventually leave $\textbf{W}$ permanently. When the dynamics within a scattering region is chaotic, we say that we have a \emph{chaotic scattering}. A typical signature for chaotic scattering is that characteristic quantities associated with the particles, or trajectories, after being scattered are sensitive to their initial conditions before entering the scattering region.

Chaotic scattering is a very active research topic in dynamical systems. It has been found to describe a very broad class of dynamical processes. To mention a few cases, one may consider the dynamics of active
chaos, i.e. chemical reagents or biological particles being advected
by the flow and at the same time undergoing intrinsic transformations,
such as biological reproduction or chemical reactions. A remarkable
example is found in the study of plankton population in the oceans~\cite{MoG04, TMG05}. 
One may also consider the dynamics of blood flow and investigate the
role of the chaotic advection in the residence time of platelets and
its consequence for the deposit and blocking of blood vessels~\cite{SGM09}. A further example may be taken from the consideration of scattering particles in celestial dynamics~\cite{saturn}, among many other dynamical phenomena. 

Because scattering dynamics are so closely related to scattering of particles in physical systems, we commonly refer to the dynamics of initial conditions in a region of the phase space as dynamical of  particles whose dynamics started in such region. From the dynamics point of view, chaotic scattering is characterised by the presence of a chaotic non-attracting set containing periodic orbits of arbitrarily large periods as well as aperiodic orbits distributed on a fractal geometric structure in the phase space --- the \emph{chaotic saddle}~\cite{Ott02}. In analogy to more general dynamics, chaotic scattering can be hyperbolic and non-hyperbolic.
In the former case, all orbits in the chaotic saddle are unstable, and a randomly
chosen initial condition has full probability of escaping within finite
time. Hyperbolic scattering is also associated with an exponential
decay of probability for particles to be found in the scattering region after a given time $t$. That is, if we have a number $N_0$ of randomly chosen initial conditions within the scattering region, the number of particles $N(t)$ in that region decays exponentially on time and consequently
\begin{displaymath}
P(t)\propto e^{-\kappa t},
\end{displaymath}
where $P(t) = \lim_{N(0) \to \infty} N(t)/N(0)$ is the probability for particles to remain in the interaction region after time $t$~\footnote{In the case of discrete time dynamical systems, $t \equiv n$, the number of iterations.}, and $\kappa$ a constant~\cite{Ott02}.
Non-hyperbolic scattering, on the other hand, is characterised by the presence of Kolmogorov-Arnold-Moser (KAM) islands in phase space. These islands surround marginally stable periodic orbits.  Orbits from
the outside can spend a long time in the vicinity of the KAM islands
before escaping, and this ``stickiness'' effect causes a slower escape
dynamics than that found in hyperbolic systems: non-hyperbolic escape
is characterised by a power law distribution of probability, 
\begin{displaymath}
P(t)\propto t^{-\alpha},
\end{displaymath}
for large enough $t$ \cite{L-F-OttPRL91, MotterPRE01, MotterPRE03}. An important characteristic is that due to the invariance of the islands and the area preserving property of the dynamics, if initial 
conditions are chosen inside an island, orbits are expected to be 
trapped there forever~\cite{Ott02}.

The above discussion refers to completely deterministic
dynamics. However, since most phenomena in nature are always subjected
to small fluctuations from their surrounding environment, a central
question is: \textbf{how does the dynamics of scattering systems
change in the presence of random perturbations?}

Here we focus on random perturbations of the parameters defining the system.  From the Dynamical Systems perspective, this situation can be formalised by the
concept of \emph{random maps}, which we will apply throughout this paper,
\begin{displaymath}
x_{n+1} = f_{n}(x_{n}),
\end{displaymath}
where we randomly choose slightly different maps $f_{n}$ for each
iteration $n$ (see Eq.~\ref{eq.map} below). It is known that such dynamics has well-defined (in the ensemble sense) values of dynamical invariants such as fractal dimensions and Lyapunov exponents~\cite{Fal86, Arn98, RGO90}. Note that we associate the choice of the map with the iteration. Therefore, all initial conditions in a given iteration are mapped by the same sequence of random maps. On the other hand, if we implement independent random noise added to each trajectory, what has been previously considered for the random noisy dynamics~\cite{PoG95, FeG97,SHS09, KrG04, KrG04b}, the situation is very different, as we shall discuss later.

Let us recall that hyperbolicity is a structurally stable
characteristic of dynamical systems, i.e. hyperbolic systems are
robust under small smooth perturbations of the system.  But on general
grounds, one expects the perturbation of a non-hyperbolic dynamical
system to result in a hyperbolic dynamics. We remark that the notion
of structural stability is related to perturbations of the whole
dynamical system and not to noisy diffusion-like perturbations of
individual trajectories~\cite{Rob99}. Although it is possible to define random
invariants in more general grounds~\cite{Arn98}, for our context it makes no
sense to talk about dynamical invariants such as the Lyapunov exponent
and fractal dimensions in the noisy dynamics in this latter case,
because trajectories will be ``smoothed-out'' in small scales and all
fine-structure dynamical structures will disappear. In the case of
perturbation of parameters, on the other hand, we have ``random maps''\cite{RGO90},
 which are known to have well-defined dynamical invariants in a
measure-theoretical sense\cite{Fal86}.

A further observation is that, since hyperbolic systems are
structurally stable, we expect that small perturbations would not
result in qualitative changes to the dynamics of hyperbolic maps. In
particular, escape should continue to be exponential. For
non-hyperbolic systems, the same can not be expected. Thus, we expect
to see qualitative changes due to the perturbations.  As an example,
we could heuristically see the KAM tori no longer as confining, since
now the perturbations can always push orbits out of the region of an island, and
conversely orbits on the outside can be pushed inside a torus by the
perturbations.  The destruction of this essential feature of
non-hyperbolic scattering leads us to hypothesise that \emph{even
arbitrarily small random perturbations to the dynamics will turn a
non-hyperbolic scattering system into a hyperbolic-like
dynamics}. Hence, we expect the probability distribution of escaping time for perturbed non-hyperbolic
systems to become exponential.

\begin{figure}[tb]
\includegraphics[width=.9\columnwidth]{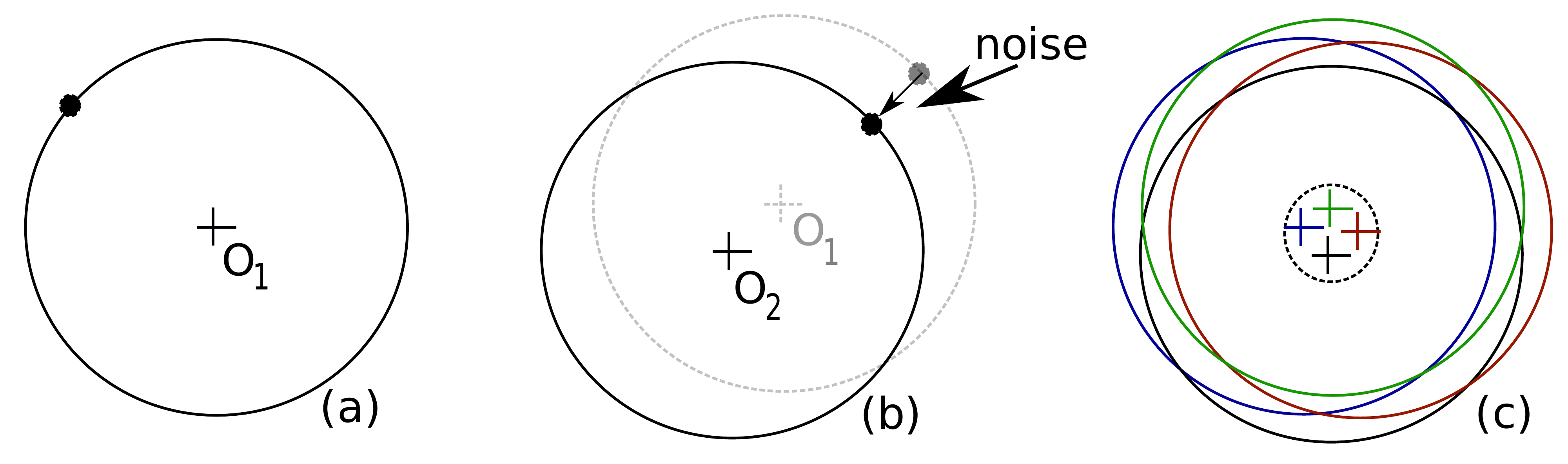}
\caption{(Colour online) Simplified representation of the orbits on
  the torus. In (a), it is represented a single torus and its
  centre. In (b), it is represented the effect of the
  perturbation. The difference between the dashed and the continuous
  line is due to the effect of the noise, that deslocate the
  centre of the torus from $O_{1}$ to $O_{2}$. In (c) we show what
  would be expected for $4$ iterations. Each continuous line represents the orbit of a particle on the torus for each
  iteration, hence different values of perturbations at each
  iteration. The centre is expected to move around $O_{1}$ at each
  iteration and for long enough time, the distribution of centres
  would fill densely the area within the dashed circle.}
\label{fig.tori}
\end{figure}

In this paper we explore and test this hypothesis using a simple
scattering map as a model system. We derive an analytical theory to
predict the scaling of the escape rate of non-hyperbolic scattering
systems under weak perturbations. The theory predicts that the
(exponential) escape rate goes to zero as the amplitude of the
perturbation decreases following a quadratic law, which we argue is
universal for all perturbed non-hyperbolic systems.  We verify this
prediction numerically, and find that it describes well the results of
our simulations.

\begin{figure}[tb]
\includegraphics[width=.9\columnwidth]{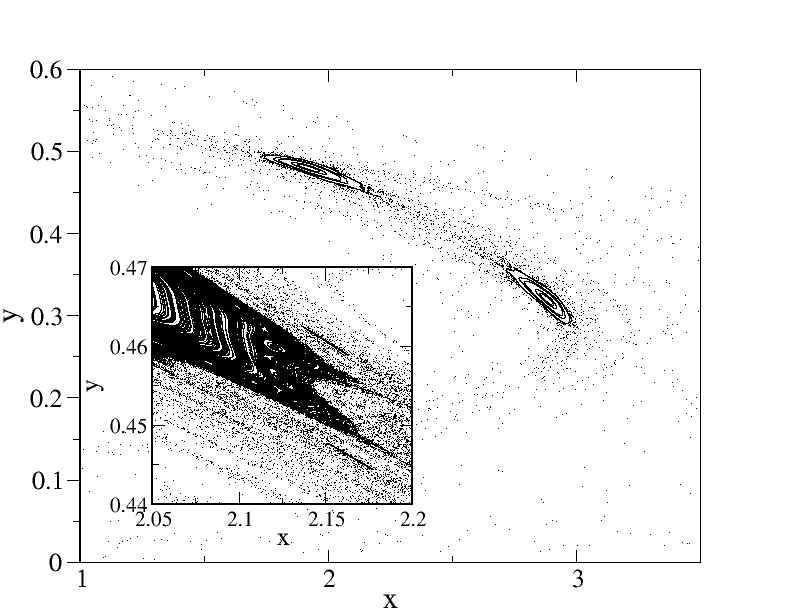}
\caption{Phase space for the map \ref{eq.map}, for $\lambda =
  6.0$. The inset shows a blowup of the region $(x,y) \in [2.05, 2.20]
  \times [0.44, 0.47]$.}
\label{fig.phase}
\end{figure}

In order to better understand the effects of random parameter
perturbations on the dynamics of non-hyperbolic Hamiltonian systems,
we develop now a simple statistical model of the motion of a single
trajectory in the presence of perturbations.  

In the absence of random perturbations, the islands act as barriers in the phase-space;
Due to the area-preserving property of such dynamics, those invariant
structures can not be transposed~\cite{Ott02}, thus trajectories trapped 
inside islands never escape. Now suppose a small
noise is added to the system's parameter. The effect of noise is to vary 
randomly the parameter from iteration to iteration, but noise amplitude 
is assumed to be small enough so that the parameter lies in the range where 
the system is non-hyperbolic. The effect of one iteration could be, say, 
to change the parameter $\lambda$ to a slightly smaller value, but still keep it
within the range such that the dynamics is non-hyperbolic. The
effect of the new parameter is then to cause a small shift in the KAM
structures compared to the value of $\lambda$ in the previous iteration.
For the next iteration another value of $\lambda$ is chosen, and the KAM
islands again move slightly in phase-space. This happens at every
iteration, and the global effect is a sort of random walk of the KAM
structures around their positions in the absence of noise. Fig~\ref{fig.tori} is
an idealised illustration of the orbit of a particle on a torus. We
represented three steps, a simple torus - Fig~\ref{fig.tori} (a)-, the effect of the
new parameter - Fig~\ref{fig.tori} (b) -, and finally, what would be ideally expected
for 4 iterations - Fig~\ref{fig.tori} (c). 

In other words, the perturbations can be imagined to cause orbits to gain
motion in the direction transversal to the tori.  The magnitude of this
transversal component of the motion is proportional to the intensity
of the perturbation (in the case of small perturbations). Since only
the transversal component can cause an orbit started within a KAM
island to escape, we focus on this component of the motion alone.

\begin{figure}[b!]
\includegraphics[width=.95\columnwidth]{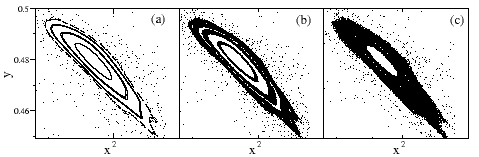}
\caption{KAM structures under the perturbed dynamics for different values
  of $\xi$. Fig~\ref{fig.phases}(a), $\xi=0$,
  Fig~\ref{fig.phases}(b), $\xi=0.0004$, and Fig~\ref{fig.phases}(c),
  $\xi=0.002$.}
\label{fig.phases}
\end{figure}

\begin{figure*}[tb]
\includegraphics[width=.95\columnwidth]{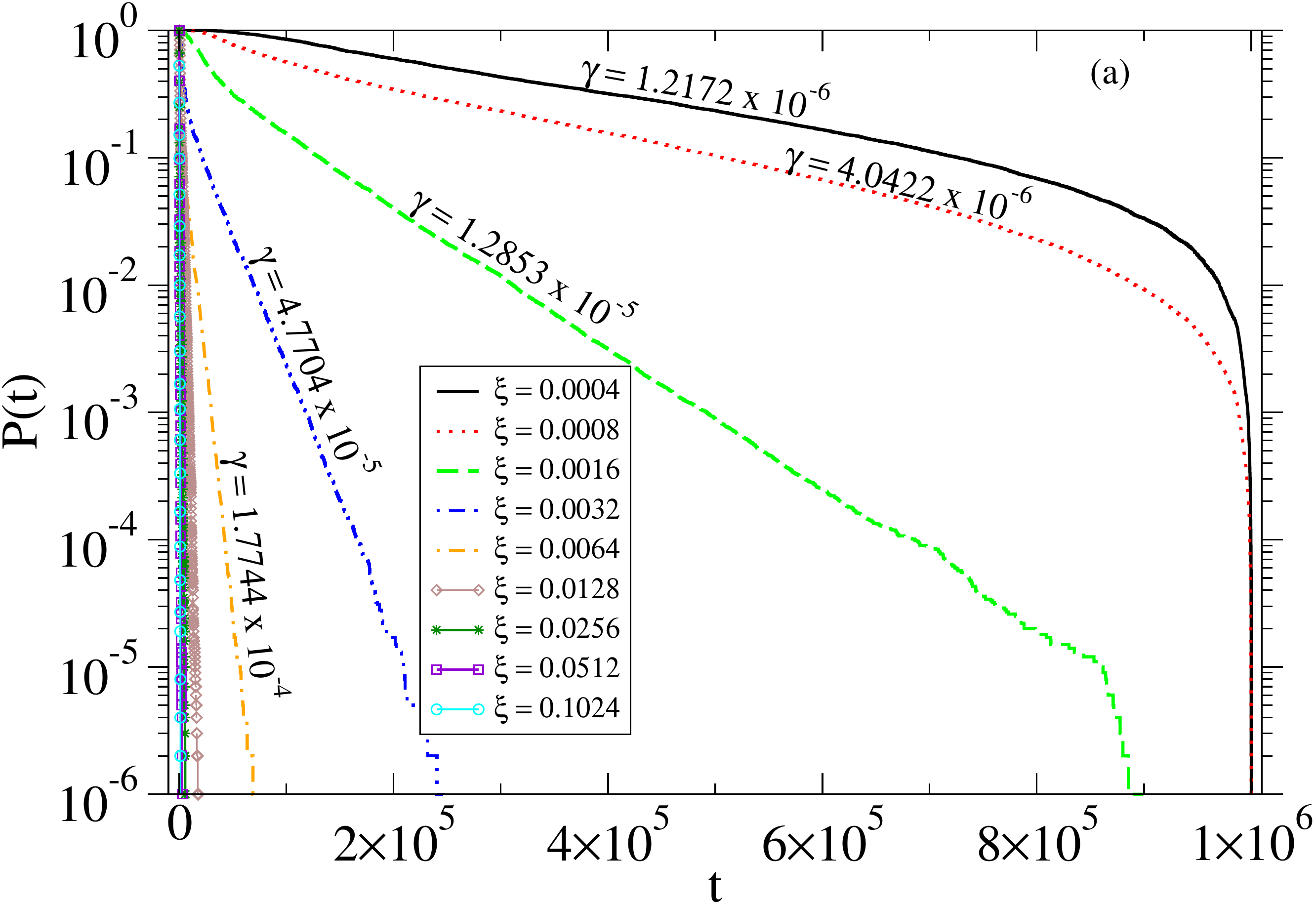}\hspace{0.1\columnwidth}
\includegraphics[width=.95\columnwidth]{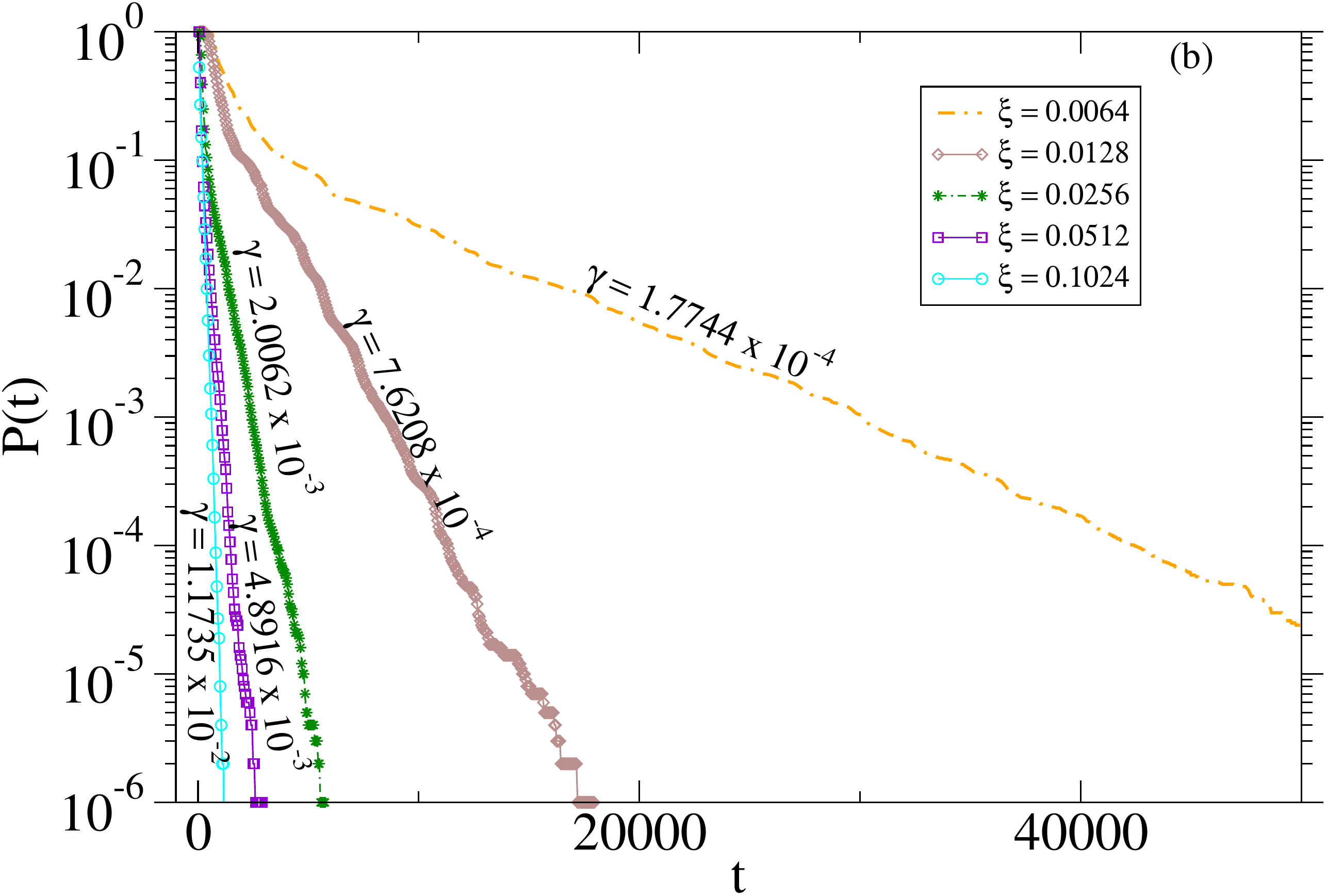}
\caption{(colour online) Probability distribution of escaping time $t$
  from the region $\boldsymbol{W} = \{|x| < 5.0, \quad and
  \quad |y| < 5.0\}$ for different values of $\xi$. Initial conditions
  were randomly chosen in the line $x \in [2.05, 2.07], y = 0.465$
  inside nest KAM structures. For each value of $\xi$, we present the
  exponent $\gamma$ that best fits the exponential region of the
  probability distribution. Fig~\ref{fig.proba}(a) shows all used
  values of $\xi$, Fig~\ref{fig.proba}(b), shows small values of $t$.}
\label{fig.proba}
\end{figure*}

The transversal motion is due entirely to the random perturbations,
and as a first approximation we model it as a one-dimensional random
walk.  The size of the step $\delta$ of the random walk is
proportional to the amplitude of the perturbation.  After $n$ steps,
the typical distance $D$ from the starting position reached by the
walker is $D \sim \sqrt{n}\delta$~\cite{Feller}.  Let $D_0$ be a
typical transversal distance a particle needs to traverse in order to
cross the last KAM surface and escape.  Thus the average time (number
of steps) $\tau$ it takes for particles to escape scale is given by
$D_{0}^{2} \sim \tau \delta^{2}$.  So $\tau$ scales as $\tau \sim
\delta^{-2}$.  The conclusion is that our simple model predicts an
exponential decay of particles, with a decay rate $\kappa\sim \tau^{-1}$ scaling as
\begin{equation}
\kappa \sim \delta^{2}.
\label{eq.law}
\end{equation}

Note that the dynamics described by random maps here is very different from that where random noise is added to individual trajectories. In the latter case, an analysis in terms of stochastic stability is more adequate, because the behaviour of, for example, two distinct orbits evolved from different initial conditions cannot be compared in terms of (Lyapunov) stability. In fact, even by following two distinct time evolutions from the same initial condition we may observe completely distinct behaviour due to the two different noise realisations, possibly not correlated. In particular, for scattering under the presence of noise~\cite{SHS09}, it has been shown that the exponential escape rate is preserved in weakly dissipative systems and that also an exponential escape rate should be expected for non dissipative ones, for noise intensity above a threshold. Indeed, because trajectories under different noise realisations may not be correlated for a certain level of noise, we expect a resultant random behaviour to be dominant, and therefore a distribution of random values of escaping time. Hence, an exponential distribution naturally follows. This scenario has already been pictured in~\cite{SHS09} by considering that above the critical level of noise, an effective random redistribution of particles results in a blurring of the fine structure of the KAM tori. In the dynamics given by random maps considered here, on the other hand, all the trajectories are evolved under the same realisation and fine-scale structures are preserved for each map, allowing a well-defined analysis of structural properties of limit sets~\cite{Arn98}. In this case, one can, for example, evaluate the Lyapunov stability of orbits by comparing the convergence towards one another or the divergence of different random orbits generated by the evolution of distinct initial conditions, since they are subject to the same sequence of perturbations. It is also possible to define consistently the fractal dimension of invariant sets under the iteration of the random maps~\cite{Fal86}, unlike in the case of independent realizations of noise for different particles. Intuitively, we can think of such a system as perturbations of the dynamics itself, for example, a set of particles that evolve subject to a field that undergoes random fluctuations. Therefore, all particles are thought to evolve under the same sequence of dynamical laws, rather than each particle being subject to random uncorrelated kicks. This key characteristic is what allows us to consider our model based on the random walk of the tori and develop our approach in order to derive the power law described by equation~\ref{eq.law}. As we shall see in what follows, it also allows us to consider singular sets of initial conditions, which we will use to compute the fractal dimension. Furthermore, we remark again that the initial conditions used here are always chosen inside the original KAM structures. For this set of particles, if no random perturbation is considered, there should not have any escape at all due to the are preserving property of our map. Simply the fact that such escape indeed happens is already by itself a novelty characteristic introduced by the perturbations of the dynamics and explained by our theory.

In order to test our theory, we numerically obtained the probability
distribution $P(t)$ that particles have not escaped at time $t$ from a
given region $\boldsymbol{W}$ of the phase-space under the dynamics of
an area-preserving non-hyperbolic map. We have chosen the map
\begin{equation}
\label{eq.map}
\begin{split}
x_{n+1} &= \lambda_{n}  [x_{n} - (x_{n} + y_{n})^{2}/4]\\
y_{n+1} &= \frac{1}{\lambda_{n}}[y_{n} + (x_{n} + y_{n})^{2}/4],
\end{split}
\end{equation}
which is non-hyperbolic for $\lambda \lesssim
6.5$~\cite{L-F-OttPRL91}. We define $\boldsymbol{W} = \{|x| < 5.0, |y|
< 5.0\}$. The initial conditions were randomly chosen with uniform
probability in the line $x \in [2.05, 2.07], y = 0.465$. For this
interval, the particles start their trajectories inside a KAM
structure (See Fig~\ref{fig.phase}).
Because we are interested in perturbing the system, we chose
$\lambda^{*} = 6.0$. Then, for each iteration $n$ we randomly chose a
perturbation $|\varepsilon_{n}|$ in the interval $|\varepsilon_{n}| <
\xi$, where $\xi$ is the amplitude of the perturbation. Indeed, for
the simulations we observe, as expected, the scenario pictured in
Fig.~\ref{fig.tori}. We show in Fig.~\ref{fig.phases} a portion of the
phase space containing KAM islands for different values of
perturbations.

We show in Fig~\ref{fig.proba} the probability distribution $P(t)$ for
different values of $\xi$.  We clearly see that the escape follows an
exponential law in a broad range of times, for various
perturbation amplitudes.  For each value of $\xi$, we show the
exponent $\gamma$ that best fits the exponential law $P(t)\sim
e^{-\gamma t}$ in the range of times for which the probability
distribution is exponential.

We also observe that for small values of the perturbation amplitude,
the stickiness plays an important role on the escaping time
distribution of particles that leave the non-attracting chaotic
set. This is because even after a particle escapes from a KAM
island, it may be trapped for long times by Cantori. Under small
amplitude of perturbations, this causes a slower escape than that
predicted by our model. This signature of the stickiness is observed 
in the tail of the distribution shown in Fig.~\ref{fig.proba} as a \textit{cutoff}
 for long times in the exponential escape of particles. The
exponential probability distribution of escape time agrees with our
hypothesis that the dynamics of the perturbed system is hyperbolic-like in
the presence of the random perturbations.
\begin{figure}[b]
\includegraphics[width=.95\columnwidth]{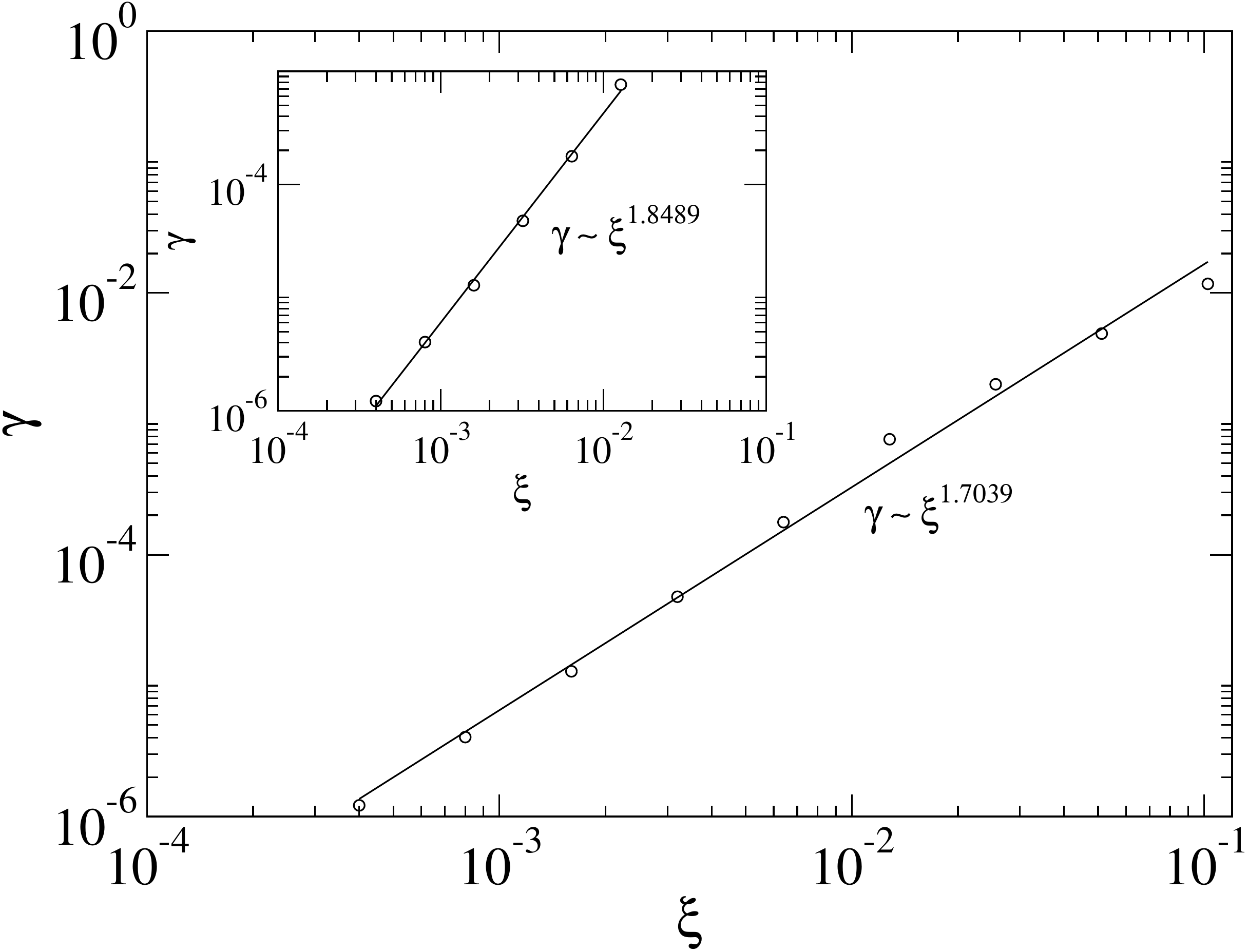}
\caption{Different values of exponent $\gamma$ as a function of $\xi$
  from Fig.~\ref{fig.proba}, as well as the power law that best fits
  the distribution. In the inset, we consider only values of $\xi <
  0.02$.}
\label{fig.expo}
\end{figure}

Using our random walk escaping model, from Eq.~\ref{eq.law} and
Fig.~\ref{fig.expo}, we identify $\kappa \equiv \gamma$, and $\delta
\equiv \xi$. In order to construct our model, we have not taken into
account the effect of stickiness. Therefore, it is reasonable to expect that the obtained exponent
should actually be smaller than $2$. We show in Fig.~\ref{fig.expo}
the dependence of $\gamma$ on the amplitude of the perturbation $\xi$,
for the map given by Eq. (\ref{eq.map}).  First of all, we notice that
$\gamma$ indeed follows a power-law, as our model predicts.  We also
obtained a good agreement with the square law predicted by
Eq.~\ref{eq.law}.

Another fundamental characteristic, differentiating non-hyperbolic
chaotic scattering from the hyperbolic case, is the fractal dimension
of singular sets. One expects that choosing initial conditions from a
line in $\boldsymbol{W}$, the set of particles that remain in
$\boldsymbol{W}$ after a given time $T_{0}$ should form a Cantor set
with fractal dimension $d<1$ in the case of hyperbolic dynamics. It is
due to the exponential decay of probability, characteristic of hyperbolic
systems. On the other hand, the algebraic decay of non-hyperbolic
scattering leads to the maximal value of the fractal dimension,
$d=1$~\cite{L-F-OttPRL91}. However, it is well known that in practice
one would need in most cases to go down to unreasonably small scales
to see the actual $d=1$ prediction in a non-hyperbolic scattering system.
For scales that are physically meaningful, in most cases one finds
that the chaotic saddle and its associated sets (stable and unstable
manifolds, etc.) have a position-dependent \emph{effective dimension}
\cite{MoG04, MotterPRE03}, which is lower than $1$.
\begin{figure}[tb]
\includegraphics[width=1\columnwidth]{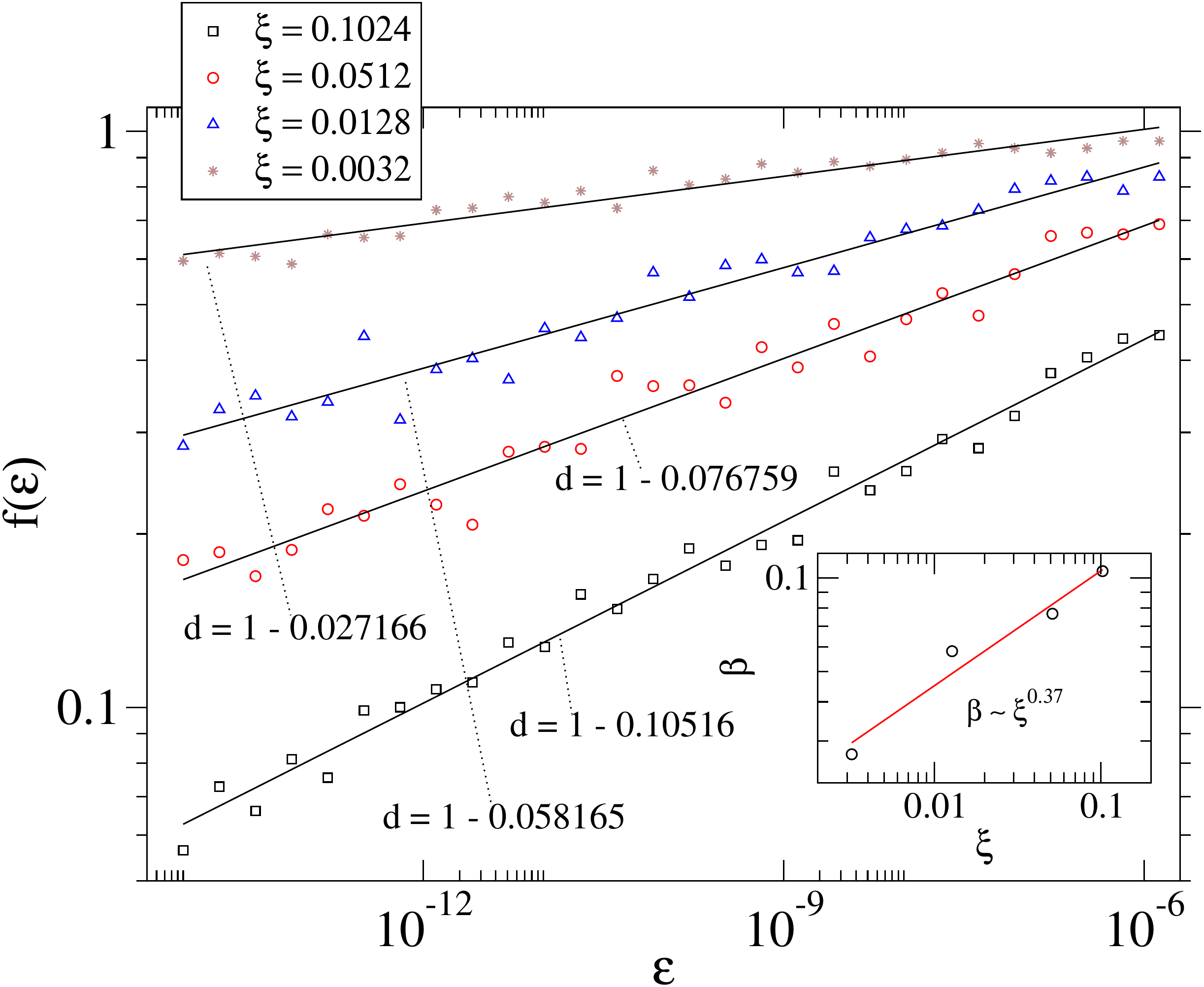}
\caption{(colour online) Estimated fractal dimension of the $T$, for
  different values of $\xi$. We notice that, when the amplitude of the
  perturbation is decreased, the dynamics approach the non-hyperbolic
  limit, and so does the estimated values of fractal dimension. The inset 
shows the value of $\beta$ as a function of $\xi$.}
\label{fig.frac}
\end{figure}

As an example of scattering function, we consider the time-delay
$T(x)$. The idea is to calculate the fraction $f(\varepsilon)$ of
pairs of initial conditions started in random positions of the phase
space with distance $\varepsilon$ of each other whose trajectories
diverge within the scattering region, and whose orbits end up escaping
through completely different routes. In order to compute that, we
chose initial conditions inside a KAM structure.  We chose $y_{0}=
0.465$, and different values of $x_{0}$ were randomly chosen
to belong to the interval $[2.05, 2.15]$. For a fixed value of
uncertainty $\varepsilon$ we chose $x_{0}$ and compute $|T(x_{0}) -
T(x_{0} + \varepsilon)|$. If $|T(x_{0}) - T(x_{0} + \varepsilon)| > 2$
we say that $x_{0}$ is $\varepsilon$-uncertain. Dividing the number of
$\varepsilon$-uncertain points by the total number of initial
conditions we obtain $f(\varepsilon)$, the uncertain fraction. It can
be shown that for chaotic scattering systems $f(\varepsilon)$ scales
as a power law $\varepsilon^\beta$, and that the box-counting fractal
dimension $d$ is given by $d=1-\beta$~\cite{grebogi83}.  We expect
that the smaller the amplitude of noise, the closer the $d$ is to $1$,
which would correspond to the non-hyperbolic limit. We show in
Fig.~\ref{fig.frac} the estimated fractal dimension of the function
$T(x)$, for different values of $\xi$.
We notice that when the amplitude of the perturbation is decreased,
the dynamics approach the non-hyperbolic limit, and so do the
estimated values of fractal dimension, which approach 1 as
$\xi\rightarrow 0$. Furthermore, we notice that $\beta$ scales as a power 
law with the amplitude of noise, i.e. $\beta \sim \xi^{0.37}$.

In conclusion, we have shown that random small perturbation of
non-hyperbolic maps leads to hyperbolic behaviour. We have presented a
random walk model to explain the escape of particles from inside
invariant KAM islands. Such particles are expected to be trapped there
forever in the absence of perturbation. When the perturbation is
present, we show that, not only particles are able to escape, but also
they have hyperbolic behaviour, with a universal quadratic power law 
relating the exponential decay to the amplitude of the perturbation. 
We further investigated the hyperbolic behaviour estimating the 
fractal dimension of the set of particles that remain inside the islands 
for a given time. We show that indeed the fractal dimension is in 
agreement to what is expected for a hyperbolic system. 

We call the attention of the reader to an important issue on
scattering dynamics clarified by our results.  Note that despite the
generality of scattering phenomena, mostly of the modelling of real
processes has assumed a simpler chaotic dynamics, that is, given by
\textit{hyperbolic structures}. Nevertheless, it is widely argued that
most of the flows experimentally significant are non-hyperbolic; see
references therein~\cite{MoG04}. In active dynamics, or blood flow, for
example, the modelling describe systems where we have trapped
vortices. Therefore, in the absence of perturbations the particles
would not leave the trapping regions; This would correspond to
plankton not being able to leave a given region in the sea, or blood
particles being trapped indefinitely in vortices, which is in contrast
with most observations. As a consequence, although the dynamics is
described by non-hyperbolic scattering, the escape is in many cases
assumed to follow an exponential law, without any further explanation~\cite{TMG05, SGM09}. 
Since it is reasonable to assume that these flows naturally
experience random perturbations, our theory provides an important
bridge to allow one to encompass natural phenomena under the
non-hyperbolic scattering framework.

Apart from striking implications to the above mentioned dynamics, our result may 
play a fundamental role on the understanding of how sources, and escaping rates of particles act on
the dynamics of Saturn Rings~\cite{saturn}. The rings are thought to be marginally
stable periodic orbits, and the gaps to be rational tori or made up by
commensurable frequencies. They are subjected to gravitational
perturbation and perturbations of the electromagnetic field, due to
solar storms, solar winds, magnetic storms, etc.~\cite{saturn}, or due
to the dynamic of ring current around Saturn~\cite{saturn}. Such
random perturbations would be represented by our perturbation on the
parameter $\lambda$ in Eq.~\ref{eq.map}. Hence, if measured, we would
expect an exponential-like escape of particles. Therefore our results
may shed some light onto the understanding of the escape of small
particles like in the Saturn's E ring~\cite{saturn}.


\end{document}